\documentclass[aps,prl,twocolumn,superscriptaddress,showpacs]{revtex4}
\usepackage[T1]{fontenc} \usepackage[latin1]{inputenc}
\usepackage{times,graphicx} \makeatletter

\begin{document}

\title{Temperature dependent \emph{d-d} excitations in manganites
  probed by resonant inelastic x-ray scattering}

\author{S. Grenier} \affiliation{Department of Physics and Astronomy,
Rutgers University, Piscataway, New Jersey 08854, USA}
\affiliation{Department of Physics, Brookhaven National Laboratory,
Upton, New York 11973, USA}

\author{J. P. Hill} \affiliation{Department of Physics, Brookhaven
National Laboratory, Upton, New York 11973, USA}

\author{V. Kiryukhin} \affiliation{Department of Physics and
Astronomy, Rutgers University, Piscataway, New Jersey 08854, USA}

\author{W. Ku} \affiliation{Department of Physics, Brookhaven
National Laboratory, Upton, New York 11973, USA}

\author{Y.-J. Kim} \affiliation{Department of Physics, Brookhaven
National Laboratory, Upton, New York 11973, USA}

\author{K. J. Thomas} \affiliation{Department of Physics, Brookhaven
National Laboratory, Upton, New York 11973, USA}

\author{S-W. Cheong} \affiliation{Department of Physics and Astronomy,
Rutgers University, Piscataway, New Jersey 08854, USA}

\author{Y. Tokura} \affiliation{Joint Research Center for Atom
Technology (JRCAT), Tsukuba 305-0046, Japan}

\author{Y. Tomioka} \affiliation{Joint Research Center for Atom
Technology (JRCAT), Tsukuba 305-0046, Japan}

\author{D. Casa} \affiliation{CMC-CAT, Advanced Photon Source, Argonne
National Laboratory, Argonne, Illinois 60439, USA}

\author{T. Gog} \affiliation{CMC-CAT, Advanced Photon Source, Argonne
National Laboratory, Argonne, Illinois 60439, USA}

\date{\today}

\begin{abstract}
We report the observation of temperature dependent electronic
excitations in various manganites utilizing resonant inelastic x-ray
scattering (RIXS) at the Mn \emph{K}-edge. Excitations were observed
between 1.5 and 16 eV with temperature dependence found as high as 10
eV. The change in spectral weight between 1.5 and 5 eV was found to be
related to the magnetic order and independent of the conductivity. On
the basis of LDA+\emph{U} and Wannier function calculations, this
dependence is associated with intersite \emph{d-d}
excitations. Finally, the connection between the RIXS cross-section
and the loss function is addressed.
\end{abstract}


\maketitle

Manganites of the form RE$_{1-x}$AE$_x$MnO$_3$ where RE is a trivalent
rare earth and AE a divalent alkali earth, exhibit a diverse range of
magnetic and electronic phases. The recent theoretical and
experimental studies have focussed on identifying the electronic
ground states and the potential role of phase inhomogeneities
\cite{SalamonRMP,dagotto01}. However, details of the origin and nature
of the electronic order remain elusive. What is known is that the
various phases are stabilized through cooperative and competing
interactions involving the spin, orbital, charge and electron-lattice
degrees of freedom of the states derived from the Mn \emph{3d} and the
O \emph{2p} bands.

Current models frequently integrate out the oxygen degrees of freedom
and parameterize the behavior of the Mn \emph{3d} orbitals with terms
such as the hopping amplitude between neighboring Mn sites, the
on-site Coulomb repulsion ($U$) and the Hund's coupling ($J_H$), each
of which are on the order of several eV.  Experimental measurements of
the excitation spectra up to these energies can thus play a key role
in the understanding of these systems - in particular such
measurements provide far more stringent tests of the various
theoretical approaches than do ground state measurements. Central to
such efforts will be understanding how the excitation spectrum relates
to the various magnetic and electronic orderings.

In this paper, we report resonant inelastic x-ray scattering studies
of the electronic excitation spectrum in a number of manganites, for a
range of ground states. We find that in all samples, the excitation
spectra show systematic temperature dependencies associated with the
magnetic ordering at energy scales up to 10 eV. The integrated
spectral weight in the 1 eV to 5 eV energy range increases on entering
ferromagnetic phases, decreases for antiferromagnetic spin alignment,
and is unchanged through metal-insulator transitions for which the
magnetic ground state is unchanged. On the basis of time-dependent
density functional calculations, we argue that this temperature
dependence arises from intersite \emph{d-d} excitations which are
suppressed or enhanced for antiferromagnetic or ferromagnetic
nearest-neighbor spin correlations, respectively. These results both
point to the sensitivity of RIXS to magnetic order and to the need for
realistic calculations of the \emph{q}-dependent dielectric function.

Inelastic x-ray scattering (IXS), like optical measurements, is a
photon in - photon out probe of electronic excitations over energy
scales up to and above the charge-transfer gap. Both techniques have
recently been applied to the manganites
\cite{Okimoto97,Jung97,Quijada98,Inami03}. IXS offers the additional
advantage that one can measure dispersion of the excitations over the
entire \emph{q}-space with bulk sensitivity.  Further, by tuning the
photon energy to the absorption edge (or ``resonance'') the IXS
cross-section may be enhanced, allowing even high-Z based systems to
be studied \cite{Kao96,schulke01} - albeit at the price of a more
complicated cross-section. Recently, Resonant IXS (RIXS) has been
utilized to study LaMnO$_3$ \cite{Inami03} where it was argued that
the technique revealed excitations related to the ordering of
orbitals.

\begin{figure}
\includegraphics[width=.8\columnwidth]{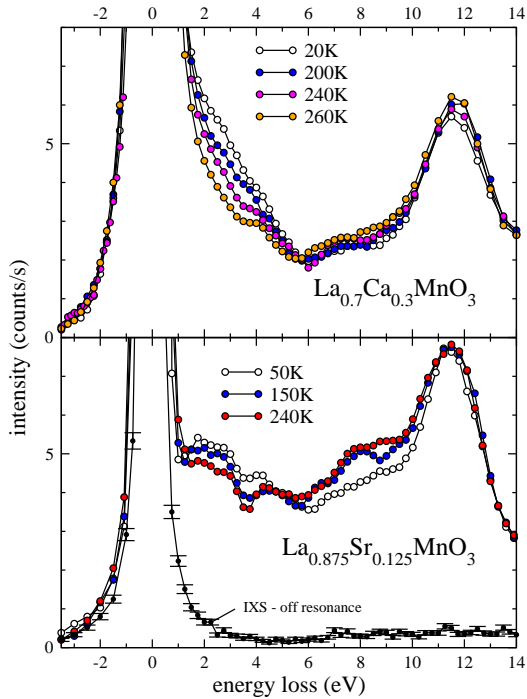}
\caption{\label{lcmo_vs_T} Energy loss spectra of LSMO.125 and
LCMO.3
 \emph{vs} temperature. Each point is averaged over three consecutive
 data points to decrease the statistical noise. The elastic line (0
 eV) was typically 3000 counts/s for LCMO.3 and 1000 counts/s for
 LSMO.125, error bars are typically .15 counts/s. An off-resonance
 spectrum is shown in the lower panel (E$_i$=6538 eV).}
\end{figure}

In the present work, samples with several magnetic, orbital, charge
and structural ground states were studied:
La$_{0.875}$Sr$_{0.125}$MnO$_3$ (hereafter refered as LSMO.125) which
has a paramagnetic semi-conductor (PMSC) to ferromagnetic metal (FM)
phase transition at 190 K and is a ferromagnetic insulator (FI) below
150 K; La$_{0.7}$Ca$_{0.3}$MnO$_3$ (LCMO.3) which is PMSC above $T_C$
= 250 K and FM below; Pr$_{0.6}$Ca$_{0.4}$MnO$_3$ (PCMO.4) which
undergoes a phase transition from PMSC to a two-dimensional
antiferromagnetic insulator (AFI) at 234 K with the onset of orbital
ordering, and becomes a three-dimensional antiferromagnetic insulator
(AFI) at $T_N=180$ K; and Nd$_{0.5}$Sr$_{0.5}$MnO$_3$ (NSMO.5) which
has a PMSC to FM transition at 250 K, and a FM to orbitally ordered
AFI transition at 150 K. All samples were grown by the floating zone
method. Their quality was checked by x-ray diffraction, resistivity
measurements and magnetization measurements using a commercial SQUID.

The experiments were performed at beamline 9IDB, CMC-CAT at the
Advanced Photon Source. Successive Si(111) and Si(311) monochromators
were used. The scattered x-rays were collected by a spherically bent
Ge(531) analyzer and focused onto a solid-state detector. The overall
resolution was 300 meV (FWHM).  The incident photons were linearly
polarized, perpendicular to the scattering plane. The incident energy,
$E_i$, was tuned to the peak of the Mn \emph{K}-edge absorption of the
respective materials - for which the RIXS intensity was maximized
(this varied from $E_i=$ 6555 to 6557 eV with increasing doping). The
data were taken at Q=(0 2.3 0), P$bnm$ settings, which minimized the
number of elastically scattered photons. Typical inelastic count rates
were between 1 and 10 counts/s. The background, as measured on the
energy gain side, was $\sim$ 0.1 count/s.  The $K_{\beta5}$ emission
line, i.e. decays from the Mn \emph{3d} band, occurs at 6534 eV,
\emph{i.e.}  out of the range of the present scans which cover the
energy $E_i+3$ to $E_i-14$ eV. Here, the inelastic scattering arises
following the recombination of the exciton comprised of the
\underline{1\textit{s}} core-hole and the 4\textit{p} photo-electron,
after it has exchanged energy with the valence electrons.

Inelastic spectra for LCMO.3 and LSMO.125 are shown in Fig.
\ref{lcmo_vs_T}. In each case, data were taken on cooling from a
paramagnetic semiconducting phase into a ferromagnetic phase:
Temperature dependence is apparent up to 10 eV. The spectra are
characterized by three distinct regions: The region up to 6 eV energy
loss, which shows an increase in spectral weight on cooling into the
respective ferromagnetic phases, a region between 6 and 10 eV which
shows little or opposite temperature dependence; and a peak at around
12 eV with no systematic temperature dependence. In what follows, we
will focus primarily on the region below 6 eV.

\begin{figure}
\includegraphics[width=1.0\columnwidth]{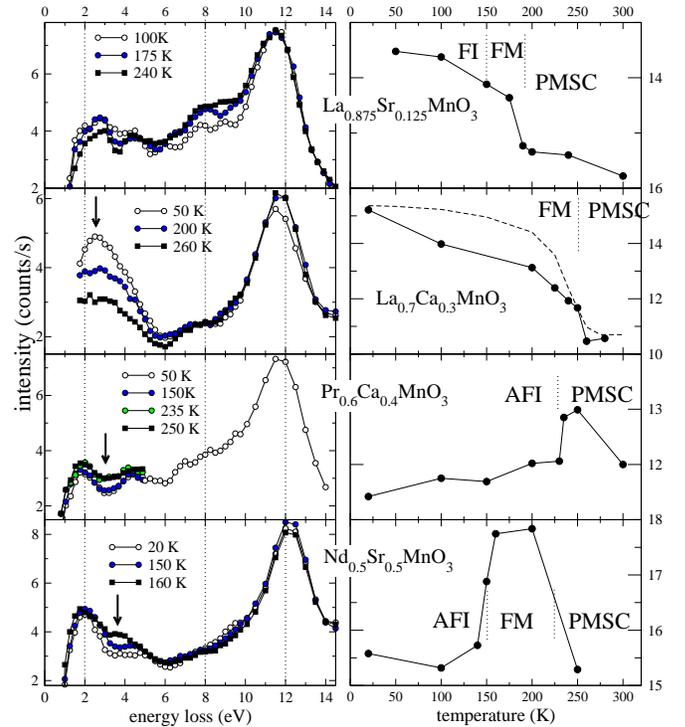}
\caption{\label{int_int} (Left) Selected energy loss spectra
\textit{vs}
  temperature after substraction of the quasi-elastic line. The arrow
  indicates the region of maximum temperature dependence. (Right)
  Integrated spectral intensity between 1.5 and 5 eV. For LCMO.3, the
  increase is shown to relate to the magnetization (dashed line,
  taken from ref.  \cite{Lynn96}) }
\end{figure}

To do so quantitatively, we must first subtract the elastic
scattering, which increases with temperature as a result of
quasi-elastic scattering from phonons. This was carried out by
assuming that the elastic line is symmetric and subtracting the
energy-gain side from the energy-loss side. This assumption was shown
to be valid by measuring the elastic scattering off-resonance
(Fig. \ref{lcmo_vs_T}). We believe this procedure gives reliable
spectra above 1.5 eV. The results for the four samples are shown in
the left panels of Fig. \ref{int_int}.

In all samples, the lowest energy feature in the subtracted spectra
occurs at $\sim$ 2 eV; at $2.75 \pm 0.1$ for LSMO.125, at $2.6\pm 0.1$
eV for LCMO.3, at $1.9\pm 0.1$ eV, for PCMO.4 and at $2.0\pm 0.1$ for
NSMO.5. This ``2 eV'' feature is reminiscent of the one observed in
manganites in optical conductivity studies \cite{Okimoto97,
Quijada98}. Note that there is an apparent trend in this feature to
move to lower energies with higher doping. Inami \emph{et al.}
reported a similar feature at $2.5\pm 0.25$ eV in LaMnO$_3$
\cite{Inami03}, which fits into this same trend. They ascribed this
excitation to an orbital excitation. The data for PCMO.4 show that
this feature is still present on warming through the orbital ordering
transition ($T_{OO}= 234$ K) and that therefore this feature cannot be
associated with long-range orbital order. More generally, our data
suggest that some contribution of the spectral weight about 2 eV is
temperature independent - see in particular the NSMO.5 data.  Thus,
the temperature dependence of the low energy spectrum does not appear
to be related to the ``2eV'' feature, but rather moves to higher
energies with higher doping (arrows in Fig. \ref{int_int}). We now
discuss this temperature dependence in detail.

In order to characterize the low energy (< 6 eV) spectra in a
model-independent way, we have simply summed the intensity between 1.5
eV and 5 eV. This integrated intensity \emph{vs} temperature is shown
as closed circles in the right panels of Fig. \ref{int_int}. These
panels illustrate the central result of this paper, namely that the
integrated spectral weight between 1.5 and 5 eV follows the
magnetization of the sample in a systematic way, and that this result
is independent of the conductivity of the ground state.

Specifically, for LSMO.125, LCMO.3 and NSMO.5 the integrated intensity
increases on cooling into a ferromagnetic metallic phase. Conversely,
the intensity is seen to drop on entering an antiferromagnetic phase
(PCMO.4 and NSMO.5). Finally, in LSMO.125 (T$_{MI}$=150 K), there is a
smooth increase in intensity as this sample is cooled through T$_{MI}$
into the ferromagnetic insulating phase. Thus, we conclude that the
change in the inelastic scattering reflects the magnetic order not the
electrical conductivity. We note that the presence of the large
quasi-elastic scattering dazzles the energy range below 1 eV where the
closing of the gap associated with the onset of metallicity occurs.

These observations raise questions as to why does the inelastic
scattering depend on the spin correlations and what is the
relationship between the RIXS spectra and any calculable response
function?  In the following, we address these issues with density
functional calculations within the LDA+\emph{U} approximation.

\begin{figure}
\includegraphics[width=.85\columnwidth]{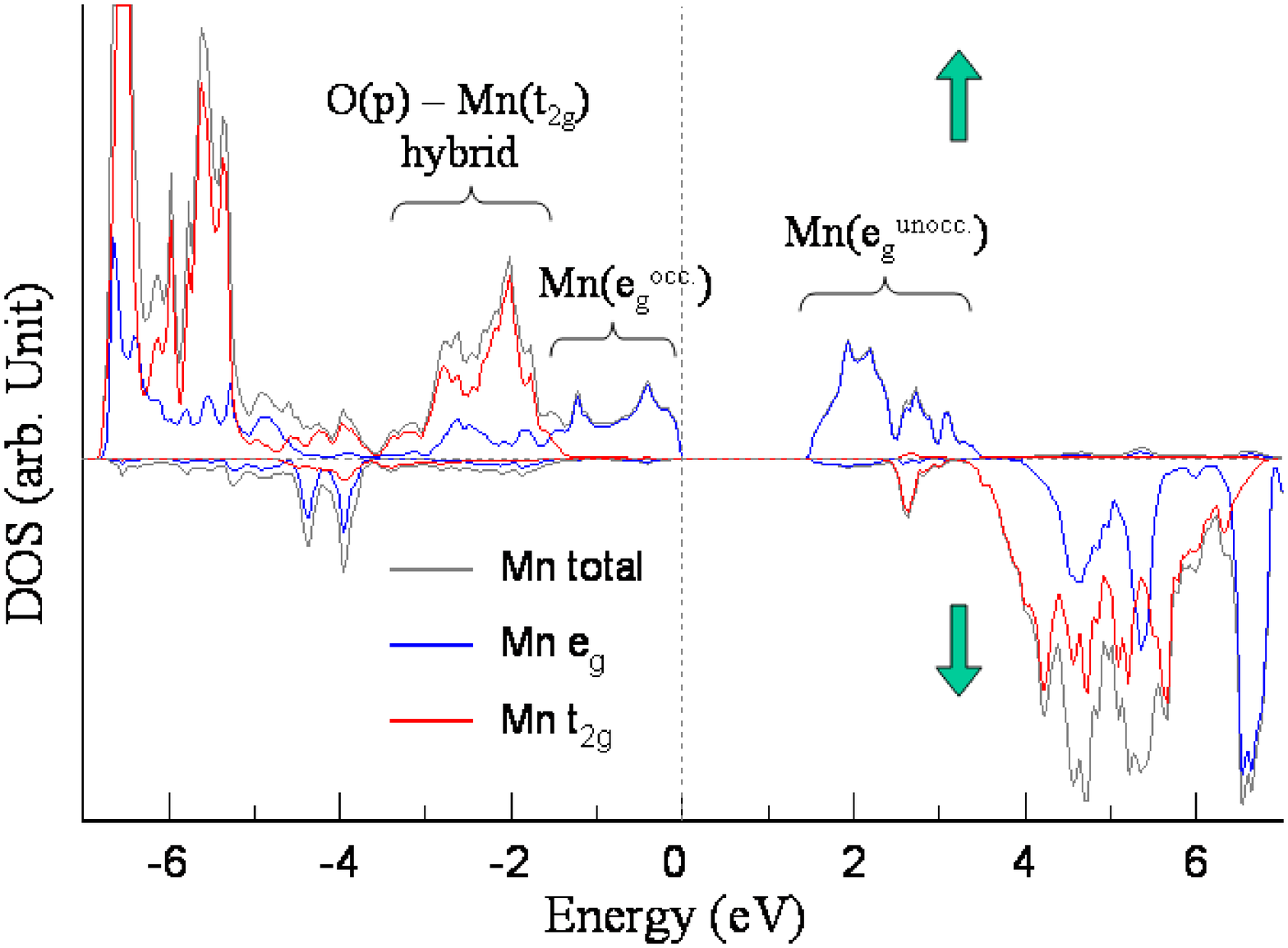}
\includegraphics[width=1.0\columnwidth]{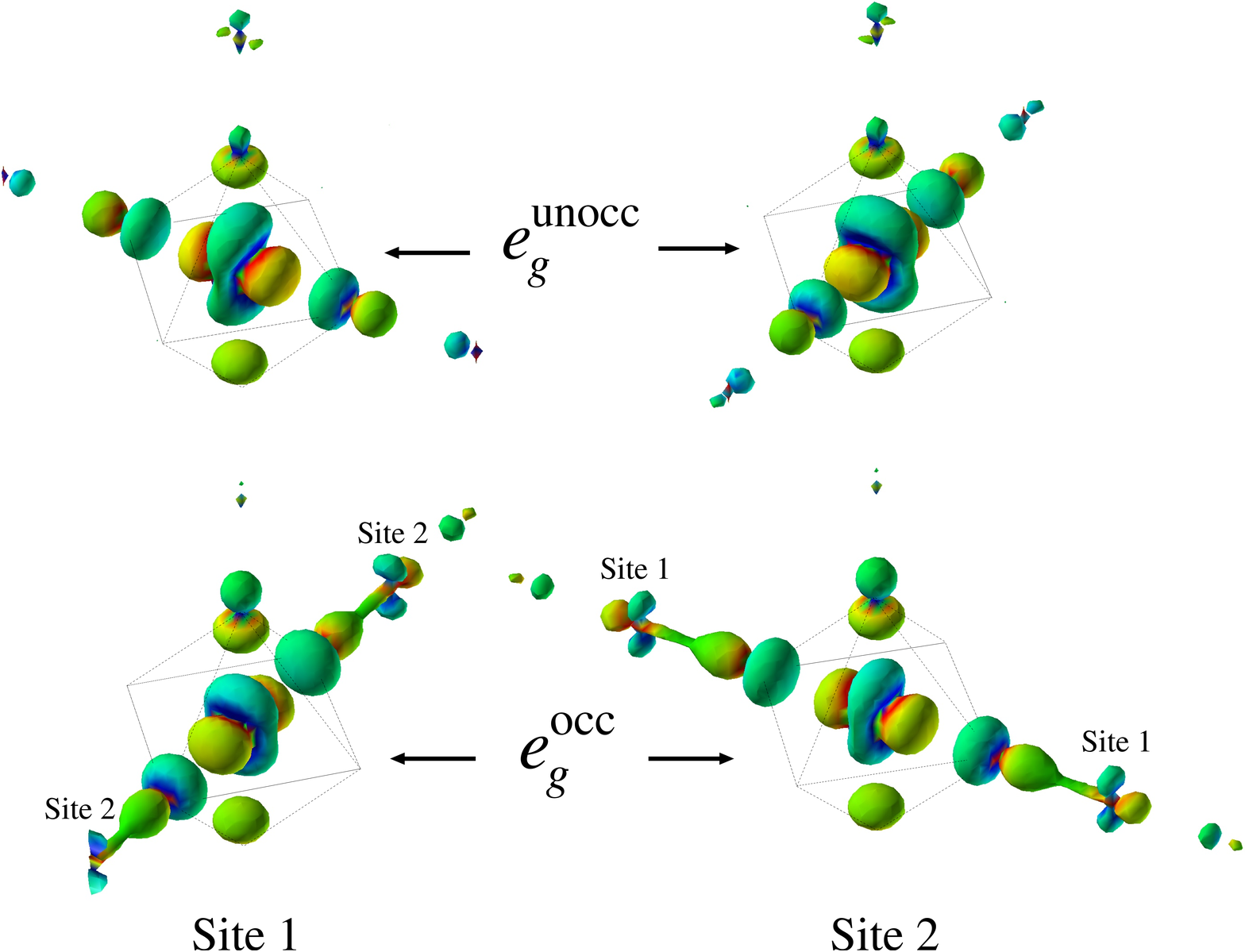}
\caption{\label{DOS} (Upper panel) Density of states on Mn atoms near
  the Fermi level for LaMnO$_3$ calculated by LDA+\emph{U}. (Lower
  panel) Wannier functions from the LDA+\emph{U} calculations for
  $e_g^{\text{occ}}$ and $e_g^{\text{unocc}}$. The oxygen octahedra
  are depicted. Site 1 and site 2 are two first neighbor Mn on the a-b
  plane (P$bnm$ setting).}
\end{figure}

We begin by calculating the electronic structure of the undoped
LaMnO$_3$ in a structure with the actual Jahn-Teller (JT) distortions
of the octahedra (but without tilting), taking $U=$ 8 eV and $J_H=$
0.88 eV (Fig. 3). We find a ground state with A-type antiferromagnetic
order and an insulating gap. The gap results from a lifting of the
degeneracy between the quarter-filled $e_g$ states that accompanies
the in-plane orbital ordering (and driven by the large on-site
repulsion \cite{Anisimov97,Volja}). These findings are consistent with
the actual ground state of LaMnO$_3$. The upper panel of
Fig. \ref{DOS} shows that the only states relevant to the low-energy
excitations are the $e_g$ states of the majority spin at the Mn
sites. The real space charge density of these states is illustrated in
the lower panel, in which the Wannier functions of both the occupied
($e_g^{\text{occ}}$) and the unoccupied ($e_g^{\text{unocc}}$) states,
either side of the Fermi level, are given for two neighboring sites
\cite{Ku:057001,Ku:167204}.  In addition to the staggered orbital
ordering of the $e_g^{\text{occ}}$ states, the unoccupied
$e_g^{\text{unocc}}$ states are spatially orthogonal to the
$e_g^{\text{occ}}$ states at the same site.

\begin{figure}
\includegraphics[width=1.0\columnwidth]{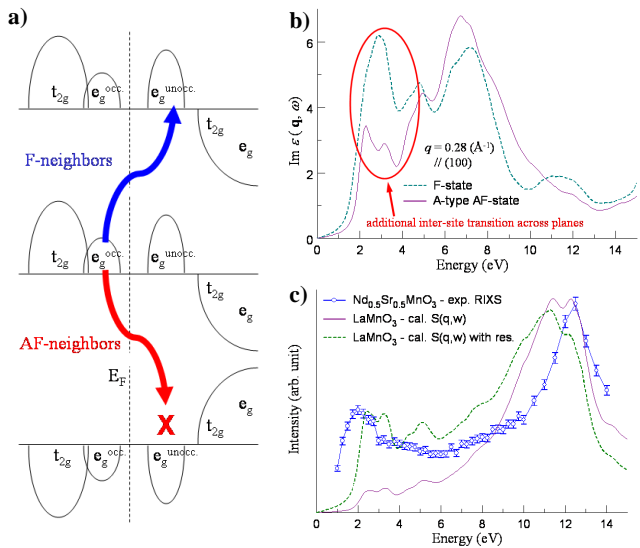}
\caption{\label{fig4} a) Schematic for the possible hopping between
  neighboring sites. For ferromagnetic neighbors, low energy
  excitations are allowed. For antiferromagnetic neighbors, such
  excitations require an improbable spin-flip and are suppressed. b)
  Imaginary part of the dielectric function for two magnetic ground
  states of LaMnO$_3$. c) Comparison between the RIXS spectra for the
  AFI NSMO.5 and the calculated S$(q,\omega)$ for AFI LaMnO$_3$, with
  and without the resonant factors (see text).}
\end{figure}

With these results in hand, the mechanism for the observed temperature
dependence of the RIXS spectra can be understood in the following way:
In the small-\emph{q} region, the low-energy features in the RIXS
spectra are dominated by \emph{inter}-site $d-d$ transitions between
Mn atoms (mediated via the hybridization with the O-\emph{2p} states
as evidenced by the Wannier functions in Fig. \ref{DOS}). Ignoring
improbable spin-flip excitations, such inter-site transitions can only
occur between neighbors with the same spin alignment, since no states
exist at low energy in the opposite spin channel. This is illustrated
schematically in Fig. \ref{fig4}a. Thus the more pairs of
ferromagnetically aligned neighbors in the system, the stronger the
low-energy spectra will be, consistent with our experiments.

This simple picture was confirmed by calculations of the dielectric
function, $\varepsilon(q,\omega)$, whose imaginary part reflects the
spectrum of allowed transitions (Fig. \ref{fig4}b)
\cite{Ku:057001,Ku3}. The calculation was performed twice with the
JT-distorted, untilted, structure: Once for the A-type AF magnetic
ordering, and once for a ferromagnetic ordering. The momentum transfer
was taken to be similar to that of the experiments, after mapping
back to the first Brillouin zone. We find that the ferromagnetic
dielectric function is doubled relatively to the AF one
(Fig. \ref{fig4}b), as is S$(q,\omega)$ at low energies (not
shown). This enhancement results from the additional charge transfer
along the c-axis, via the apical O (Fig. \ref{DOS}), which is
suppressed in the AF case. This additional contribution is of similar
size to that arising from transitions within the a-b plane because,
despite the ferromagnetic correlations of the four Mn nearest
neighbors in-plane, only two contribute due to the directional nature
of the ordered orbitals - as illustrated by the Wannier functions
(Fig. \ref{DOS}). In the actual doped systems, this difference may be
reduced for various reasons, including changes in orbital ordering,
degree of hybridization, the bandwidth of the $e_g$ states, and
screening due to metallization. However, the qualitative trend is
expected to persist and explains the present results.

Our first principles calculations also allow us to discuss the
connection between the RIXS cross-section and a calculable response
function. Several authors have argued that for \emph{K}-edge resonant
IXS, the cross-section is proportional to the IXS cross-section,
$\text{S}(q,\omega)$ (which is proportionnal to $
\text{Im}(\frac{-1}{\varepsilon})$, the loss function) multiplied by a
term accounting for the resonance
\cite{Platzman98,Abbamonte99,vdBrink03}. In Fig. 4c we show a
comparison between the NSMO.5 data (low temperature AFI) and the
calculated S$(q,\omega)$ for LaMnO$_3$ (low temperature AFI), weighted
or not by the resonant factor. It is seen that the weighted
S$(q,\omega$) closely resembles the data, clearly suggesting that
there is a connection between the two, though this requires more work
before a firm conclusion can be drawn.

Before concluding, we briefly comment on the high energy portion of
the spectra. The 8 eV region has little temperature dependence except
for LSMO.125 which exhibits a clear drop on cooling through $T_C=190$
K with a continuous decrease through the orbital ordering transition
at $T_{OO}=150$ K. In Fig. \ref{fig4}a, we see that the 8 eV region
corresponds to transitions from the majority states to the minority
states of an AF neighbor. The decrease within a ferromagnetic phase is
therefore consistent with our picture. As to the 12 eV feature, it was
observed in LaMnO$_3$ with RIXS \cite{Inami03} and was attributed to a
transition involving Mn states. Conversely, no excitation is seen
around 12 eV in the optical spectra of the manganites
\cite{Arima95,Okimoto97}. As shown in Fig. \ref{int_int}, substituting
the earth cations does not significantly change the position of this
feature, implying that their electronic states are not involved.  In
fact, it appears clearly here as a feature in S$(q,\omega)$
(Fig. \ref{fig4}c) and our calculations suggest that it is more of a
collective response of the system as a whole, \emph{i.e.} a plasmon,
as $|\varepsilon|$ reaches a low minimum.

In summary, we have measured temperature dependence in the inelastic
x-ray scattering spectra of several manganites up to 10 eV energy
loss. In the lowest energy range, we find that the excitation spectrum
does not depend on the presence or absence of orbital order, but does
correlate with the magnetic order of the sample. We calculated the DOS
for LaMnO$_3$ as well as the Wannier functions for the $e_g$ states
and describe the magnetic order dependent feature as arising from
inter-site \emph{3d}-\emph{3d} excitations. This work points to the
sensitivity of the RIXS technique to magnetic correlations in
manganites in addition to the charge excitations.

We acknowledge fruitful discussions with P. Abbamonte, V. Oudovenko,
J. Rehr, J. van den Brink and M. van Veenendaal. Use of the Advanced
Photon Source was supported by the U.S. DOE, Office of Basic Energy
Sciences, under Contract No.W-31-109-Eng-38. Brookhaven National
Laboratory is supported under DOE Contract No. DE-AC02-98CH10886.
Supports from the NSF MRSEC program, Grant No. DMR-0080008 and from
the NSF Grant No. DMR-0093143 are also acknowledged.

\bibliographystyle{apsrev}
\bibliography{/home/stephane/Papers/BiblioTex/declaration,/home/stephane/Papers/BiblioTex/rixs,/home/stephane/Papers/BiblioTex/manganites,/home/stephane/Papers/BiblioTex/rxs,/home/stephane/Papers/BiblioTex/general}
\end{document}